\documentclass[10pt,twocolumn,twoside]{elsart}

\usepackage{graphicx}
\usepackage{amsmath}

\newcommand{ \auau }{$^{197}\mathrm{Au}+^{197}\mathrm{Au}$ }

\newcommand{ \pt } {$p_T$}
\newcommand{\rts}{$\sqrt{s_{NN}}$ }

\begin{document}

\begin{frontmatter}

\title{Resonance decay effects on anisotropy parameters}

\author{X. Dong$^{a,b}$, S. Esumi$^{c}$, P. Sorensen$^{b}$,
  N. Xu$^{b}$, Z. Xu$^{d}$} \address{$^a$Department of Modern Physics,
  University of Science and Technology of China, Hefei, 230026,
  China\\ $^b$Nuclear Science Division, Lawrence Berkeley National
  Laboratory, Berkeley, CA 94720, USA\\ $^c$Institute of Physics,
  University of Tsukuba, Tsukuba, Ibaraki 305, Japan\\ $^d$Physics
  Division, Brookhaven National Laboratory, Upton, NY 11973, USA}

\begin{abstract} We present the elliptic flow $v_2$ of pions produced from
  resonance decays. The transverse momentum $p_T$ spectra of the
  parent particles are taken from thermal model fits and their $v_2$
  are fit under the assumption that they follow a
  number-of-constituent-quark (NCQ) scaling law expected from
  quark-coalescence models. The $v_2$ of pions from resonance particle
  decays is found to be similar to the measured pion $v_2$.  We also
  propose the measurement of electron $v_2$ as a means to extract
  open-charm $v_2$ and investigate whether a thermalized system of
  quasi-free quarks and gluons (a quark-gluon plasma) is created in
  collisions of Au nuclei at RHIC.
\vspace{1pc}
\end{abstract}


\end{frontmatter}

{\it Introduction.}---One of the surprising observations made at
RHIC is the measurement of a number-of-constituent-quark (NCQ)
dependence for both elliptic flow $v_2$ and the nuclear
modification factor $R_{CP}$ at intermediate $p_T$ ($1.5 < p_T<
5$~GeV/c)~\cite{starklv2}. Models of hadron formation by
constituent-quark coalescence provide a viable explanation for
these observations whereas expectations based on conventional
fragmentation approaches are inconsistent with the
data~\cite{voloshinqm02,friseprc68,coal}.  In coalescence models
an NCQ-scaling of $v_2$ arises as a consequence of hadrons
coalescing out of a thermal distribution of partons and reveals
the flow developed during a partonic epoch at RHIC. Pion $v_2$,
however, appears to violate NCQ-scaling.  In this paper, we study
the effect of resonance decays on pion $v_2$.  We show that when
decays are taken into account, the measured pion $v_2$ may become
consistent with the NCQ-scaling demonstrated by kaon ($K^+, K^-,
K^0_S$), proton, $\Lambda$, and $\Xi$
$v_2$~\cite{starklv2,phenixv2}.

The particle azimuthal distribution with respect to the reaction plane
at rapidity $y$ can be described by a Fourier expansion:
\begin{equation}
\frac{dN}{d\Delta\phi} \propto 1 + \Sigma_{n}2 v_{n}\cos(n\Delta\phi),
\end{equation}
where $\Delta\phi$ is the difference in azimuth angle between the
particle and the reaction plane. The first and second Fourier
coefficients, $v_1$ and $v_2$, historically are called directed
and elliptic flow, respectively. All coefficients can be
calculated from the relation: $v_{n} = \langle \cos(n \Delta
\phi) \rangle$.

As the volume of the system created in an off-axis nucleus-nucleus
collision expands, its spatial anisotropy quenches. The momentum-space
anisotropies represented by the Fourier coefficients $v_n$ preserve
information about the early collision dynamics when the spatial
anisotropy was
largest~\cite{voloshinqm02,sorge97,ollitrault92,xunu2}. Since the
initial overlap region is elliptical in shape, the second harmonic
coefficient $v_2$ is the largest and most studied.

\begin{figure}[ht]
\centerline{\includegraphics[width=0.5\textwidth,
    height=0.55\textwidth] {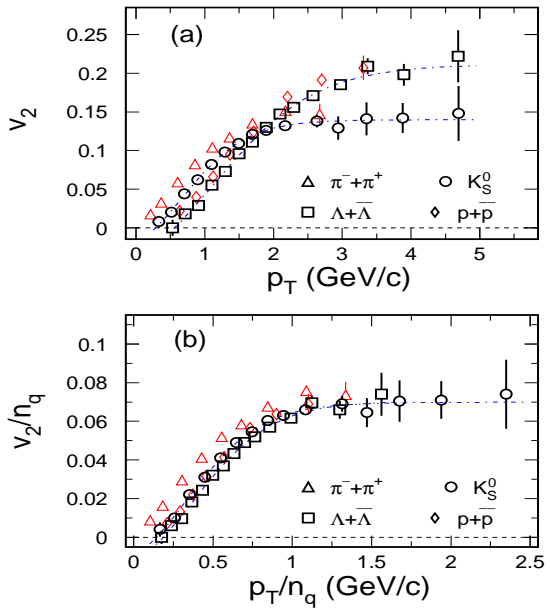}}
\vspace{-.5cm}
    \caption[]{(a) Measurements of the $p_T$ dependence of the event
      anisotropy parameters for $\pi$, $K_S^0$, $p$,
      $\Lambda$. Dot-dashed lines are the results of fits; (b)
      Number-of-constituent-quark (NCQ) scaled $v_2$. All particles
      except the pions follow the NCQ scaling. }
    \label{datav2} \end{figure}

{\it NCQ-scaling of $v_2$.}---Fig.~\ref{datav2} shows the
$\pi^++\pi^-$, $K_S^0$, $p+\overline{p}$, and
$\Lambda+\overline{\Lambda}$ $v_2$ from minimum-bias \auau
collisions at \rts = 200 GeV \cite{starklv2,phenixv2}.  In the
lower $p_T$ region ($p_T<1.0$~GeV/c), the values of $v_2$ are
lower for higher mass hadrons. Hydrodynamic calculations
\cite{pasi1} predict the observed mass dependence of
$v_2$--perhaps implying that a thermalized system has been created
in collisions at RHIC energy.  At higher $p_T$ ($p_T \ge
2$~GeV/c), the $v_2$ measurements saturate at values below the
hydrodynamic model predictions. The saturated value of $v_2$ and
the $p_T$ scale where the saturation sets in depends on the
particle-type: the baryon $v_2$ saturates at higher $p_T$ and at
larger values than that of mesons.

According to coalescence models~\cite{coal}, after scaling both
$v_2$ and $p_T$ with the number of the constituent quarks (NCQ) in
the corresponding hadron, all particles at intermediate $p_T$
should fall onto one universal curve. The NCQ-scaled $v_2$
measurements in Fig.~\ref{datav2}-(b) show that
$v_2$($p_T/n_{q}$)/$n_{q}$ for $p_T/n_{q}
> 0.6$~GeV/c is similar for all particles {\it except} pions. This
observation, coupled with the NCQ-dependence observed at
intermediate $p_T$ in the nuclear modification factor $R_{CP}$ is
evidence of hadron formation by coalescence or recombination. In
this case, $v_2$($p_T/n_{q}$)/$n_{q}$ represents a constituent
quark momentum-space anisotropy $v_2^q$ that arises as a
consequence of collectivity in a partonic stage. Based on
coalescence models, NCQ-scaling suggests the creation of a
quark-gluon plasma (QGP) with $v_2^q$ characterizing the
properties of the QGP. For this reason, understanding the source
of the discrepancy in the NCQ-scaled pion $v_2$ is imperative.

With this goal in mind, we study the effect of secondary pions
(from particle decays) on the measured pion $v_2$. We assume that
NCQ scaling is valid for all hadrons other than pions and use the
published $v_2$ measurements \cite{starklv2,phenixv2} to
parameterize $v_2$($p_T/n_{q}$)/$n_{q}$. The \pt\ distributions
are assumed to follow an exponential form with slope parameters
taken from measurements when available. We use chemical fits to
fix the relative hadron abundances~\cite{pbm,nxu01}. Since the
pion mass is much smaller than the sum of its constituent quarks
masses, direct pions are not necessarily assumed to follow the
scaling predicted from coalescence models. As such, we do not
consider direct pions, and instead choose to study the $v_2$ of
the secondary pions. Given the model uncertainties, extraction of
the direct pion $v_2$ is difficult and remains an open question.
Finally we will discuss how to extract open-charm $v_2$ based on
the decayed electrons.

{\it Simulation results.}---The $v_2$ values of the simulated
resonances are parameterized by fitting $K_S^0$ and $\Lambda$
$v_2$~\cite{starklv2} with the equation:
\begin{equation}
f_{v_2}(n) = \frac{an}{1+exp(-(p_T/n-b)/c)}-dn, \label{eq1}
\end{equation}
where $a,b,c$ and $d$ are the fit parameters, $n$ is the
constituent-quark number, and $p_T$ is in the unit of GeV/c. The
fit results are shown as dot-dashed lines in Fig. \ref{datav2},
where the fitting parameters are $a = 0.1$, $b=0.35$, $c = 0.2$
and $d=0.03$. The NCQ-scaling of $v_2$ works well for kaons,
protons and Lambdas within $0.6 \le p_T/n_q \le 1.5$ GeV/c whereas
pion $v_2$ deviates from NCQ scaling for all $p_T$.  The
parameters from chemical fits are listed in Table~\ref{tab1}.

\begin{table}[htb]
\caption{Parameters for the input resonances: the units for slope
parameters T are GeV. The fraction of the hadrons are fixed from
the measured
  abundances.}
\label{tab1}
\begin{tabular}{@{}llllc}
\hline
           & {T1 }       & {T2 }      & {T3 }      & {Fraction (\%)}\\ \hline
$\rho$     & {$0.5 $}   & {$0.4 $}  & {$0.3 $}  & {$60 \pm 10$}      \\
$\omega$   & {$0.5 $}   & {$0.4 $}  & {$0.3 $}  & {$30 \pm 10$}      \\
$K_S^0$    & {$0.3 $}   & {$0.3 $}  & {$0.3 $}  & {$6 \pm 5$}        \\
$K^*$      & {$0.5 $}   & {$0.4 $}  & {$0.3 $}  & {$2  \pm 1$}       \\
$\Delta$   & {$0.55 $}  & {$0.55 $} & {$0.55 $} & {$2 \pm 2$}        \\ \hline
\end{tabular}
\end{table}

In high-energy collisions, a large fraction of hadrons are produced
through resonance decays. This is particularly true for pions in
high-energy heavy-ion collisions. At mid-rapidity, in collisions at
RHIC, as many as 80\% of pions are from resonance decays
\cite{zbxu03}. The dominant decays are $\rho \rightarrow \pi\pi$,
$\omega \rightarrow 3\pi$, $K^*(892) \rightarrow K\pi$, $K_S^0
\rightarrow \pi\pi$ and $\Delta \rightarrow N\pi$. With such a
potentially large fraction of pions arising from decays, accounting
for their effect on the observed pion $v_2$ is very important.

\begin{figure}[ht]
\centerline{\includegraphics[width=0.45\textwidth,
    height=0.5\textwidth] {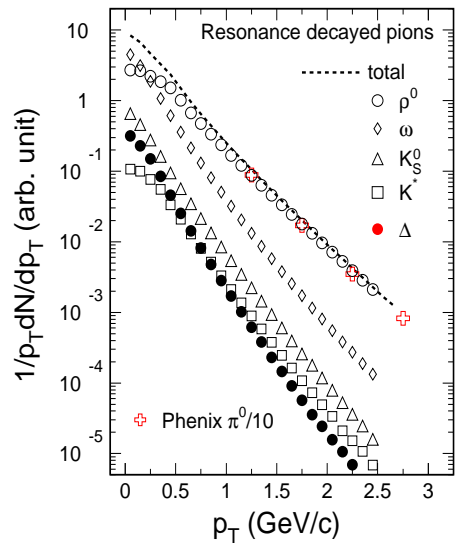}}
\vspace{-.75cm}
    \caption[]{Resonance decayed pion distributions. The summed
      spectrum is shown as a dashed-line. In this simulation, a slope
      parameter of T = 0.4 GeV is used for $\rho, \omega$ and
      $K^*$. For $K_S^0$ and $\Delta$ the respective slope parameters
      T = 0.3 GeV and 0.55 GeV are used. The relative fraction of the
      hadrons are listed in Table \ref{tab1}. For comparison, PHENIX
      $\pi^0$ results are shown as open-crosses.}
    \label{spectra} \end{figure}

The $p_T$ distributions of pions from resonance decays are shown
in Fig. \ref{spectra}. Many pions at low $p_T$ are generated from
resonances at larger $p_T$ where presumably the parent particle
$v_2$ is larger. As a result the decayed pions take on a
relatively large $v_2$ value.  The decays from the $\rho$- and
$\omega$-mesons dominate the secondary pion $p_T$ spectrum. In
this plot, a slope parameter of T = 0.4 GeV is used for the $\rho$
distributions. In peripheral collisions, the STAR measured slope
parameter is 319$\pm4$(stat.)$\pm 32$ (syst.) MeV \cite{starho}.
The simulated results are in a good agreement with the PHENIX
$\pi^0$ data from minimum bias \auau collisions~\cite{phenixpi0}.

\begin{figure}[ht]
\centerline{\includegraphics[width=0.45\textwidth]{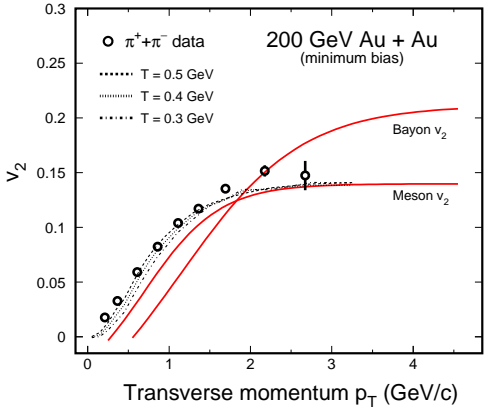}}
\vspace{-.75cm}
    \caption[]{The measured pion $v_2$ (symbols) is compared to the
      simulated $v_2$ for pions from resonance decays (dashed
      lines). The assumed $v_2$ of mesons and baryons are represented
      by the solid and dot-dashed lines, respectively.}
    \label{simv2} \end{figure}

In Fig. \ref{simv2}, the $v_2$ values for the simulated decay pions
are shown as dashed-lines. The resonances included in this study are
the $\rho$, $\omega$, $K^*$, $K_S^0$ and $\Delta$. The decay $\rho
\rightarrow \pi\pi$ with a 100\% branching ratio dominates the
production of secondary-pions. The simulated resonance particles are
restricted to mid-rapidity $|y|< 0.5$. Increasing the rapidity window
does not change the results. The pion $v_2$ in the region \pt\ $\le
1.5$ GeV/c is sensitive to the shape of the $\rho$ \pt\
spectrum. Dashed-, dotted-, and dot-dashed-lines correspond to the
$v_2$ results from the different slope parameters listed in
Table~\ref{tab1}. For the smaller slope parameter T = 300 MeV, the
decayed pion $v_2$ is below the data, leaving room for other
contributions \cite{Ko04}.

The effects of resonance decays on proton and kaon $v_2$ were also
investigated. As in Ref.~\cite{Ko04}, they were found to increase the
observed $v_2$ above the assumed NCQ-scaled input $v_2$ but by a much
smaller degree than that of pions. Since the $v_2$ of kaons is used in
our fits, resonance decays can change our assumed input $v_2$. These
changes, however, will be small. In this paper we concentrate on
the pion $v_2$.

{\it D-meson $v_2$.}---NCQ scaling suggests that hadrons at
intermediate $p_T$ are formed from a thermal partonic phase
created in heavy-ion collisions at RHIC. In this system, the high
initial matter density gradient and copious interactions among
partons leads to a collective motion. The large $v_2$ values
measured for multi-strange hadrons also indicate that partonic
collectivity develops in collisions at RHIC \cite{xiom130}. The
demonstration of collectivity is necessary but not sufficient to
show that local thermal equilibrium is established. If
collectivity is developed by much heavier charm quarks, however,
it suggests that interactions between charm quarks and $u$-, $d$-,
or $s$-quarks must have been frequent enough and strong enough for
the lighter quarks to have thermalized. For this reason, the
measurement of heavy flavor (open-charm) $v_2$ can probe the
degree to which the lighter $u$-, $d$-, and $s$-quarks thermalize.

\begin{figure}[ht]
\centerline{\includegraphics[width=0.45\textwidth,
    height=0.5\textwidth] {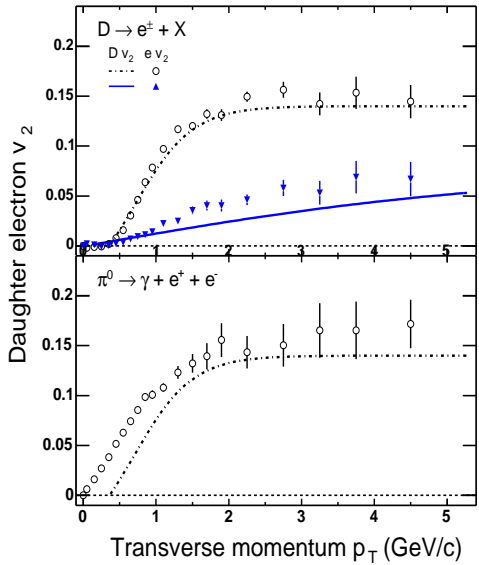}}
\vspace{-.4cm}
    \caption[]{The $v_2$ of electrons from D-meson decays (top) and
      $\pi^0$ decays (bottom). An input D-meson $v_2$ from a
      coalescence model with zero charm quark $v_2$~\cite{Lin:2003jy}
      is shown as a solid line (top). The input meson $v_2$ curve
      taken from our fit to the NCQ-scaled $K_S^0$ and
      $\Lambda+\overline{\Lambda}$ $v_2$ is shown as a dot-dashed line
      in both plots.}
    \label{dev2} \end{figure}

Since a large fraction of heavy-flavor hadrons decay through
leptonic modes, electron $v_2$ measurements may offer a convenient
way to study open-charm $v_2$. In Fig. \ref{dev2}-(top), the $v_2$
for electrons from D-meson decays is shown.  Dot-dashed lines
represent the $v_2$ of the parent meson taken from our fit to the
NCQ-scaled $K_S^0$ and $\Lambda+\overline{\Lambda}$ $v_2$. The
solid line represents a D-meson $v_2$ where D-mesons are assumed
to coalesce from light quarks with $v_2^{u,d}>0$ and charm quarks
with $v_2^c=0$~\cite{Lin:2003jy}. We simulate the D-meson $p_T$
distribution using the PYTHIA event generator \cite{pythia}. The
shape of the calculated spectrum is well represented by a
power-law function. Approximately 50 M D-meson events are used in
this simulation. For electron $p_T > 1.5$~GeV/c, and a saturated
$v_2$ at higher $p_T$ the $v_2$ of electrons from D-meson decays
becomes similar to the parent D-meson $v_2$. The decay electron
$v_2$ is found to be sensitive to changes in the assumed D-meson
$v_2$.  As such, the measurement of electron $v_2$ can distinguish
between $v_2^c\approx v_2^{u,d}$ and $v_2^c\approx 0$. The degree
of heavy flavor thermal equilibrium can be assessed by measuring
electron $v_2$ within $1 \le p_T(\mathrm{e}) \le 3$ GeV/c, a
region which corresponds to $2\ ^{<}_{\sim}\ p_T(\mathrm{D})\
^{<}_{\sim}\ 5$ GeV/c. The authors of Ref. \cite{Greco:2003vf}
come to a similar conclusion but do not evaluate background
contributions to the electron spectrum.

Neutral pion decays are the dominant source of ``background''
electrons. While the two-photon decay process,
\begin{center}$\pi^0
\xrightarrow{\sim 100\%} \gamma +\gamma \xrightarrow{\mathrm{few }\%}
e^++e^-+e^++e^-,$
\end{center}can be identified by its decay topology
\cite{starpi0}, the pion Dalitz decay can only be subtracted
statistically. In Fig.~\ref{dev2}-(b) we show the $v_2$ of electrons
from simulated pion Dalitz decays. The pion distribution can be
obtained from measurements at RHIC \cite{phenixpi0}. For these
simulations a 100\% conversion probability is assumed. In STAR,
however, the probability is closer to 5\%. The decayed electrons
predominantly have $p_T \le$ 0.5 GeV/c \cite{starpi0}.

Electrons from heavy flavor decays begin to dominate the electron
spectrum above $p_T \sim 3$~GeV/c. With knowledge of the pion yield,
D-meson yield, the pion $v_2$, and the electron $v_2$ it will be
possible to extract the D-meson $v_2$. These measurements can be made
by both the PHENIX and STAR collaborations at RHIC. Direct photon
$v_2$ can also be measured with this method.

{\it Summary.}---We have studied the effect of resonance decays on the
pion $v_2$ in Au+Au collisions at RHIC. When the $v_2$ values for
resonances are assumed to follow NCQ scaling, the pions generated in
their decays take on $v_2$ values similar to those measured at
RHIC. The dominant source of secondary pions is $\rho$ decays. We have
shown that when decays are accounted for, the measured pion $v_2$
values {\it may} become consistent with the NCQ scaling law that suggests
the development of partonic collectivity in collisions at RHIC. Model
uncertainties, however, make it difficult to extract the direct pion
$v_2$. In addition, we propose the measurement of electron $v_2$ as a
means to study open-charm $v_2$ and hence the degree of thermalization
reached at RHIC.


\par {\bf Acknowledgements}: We appreciate fruitful discussions with
V. Greco, P. Huovinen, C. Ko, H.G. Ritter, E.V. Shuryak and S.
Voloshin. This work was supported in part by the NSFC under the
Project No. 10275027 and the U.S. Department of Energy under
Contract No. DE-AC03-76SF00098.



\begin{thebibliography}{00}



\bibitem{starklv2} J.~Adams {\it et al.} [STAR Collaboration],
Phys. Rev. Lett. {\bf 92}, 052302 (2004).

\bibitem{voloshinqm02}
S.~A.~Voloshin,
Nucl. Phys. A {\bf 715}, 379 (2003).

\bibitem{friseprc68}
R.~J.~Fries, B.~Muller, C.~Nonaka and S.~A.~Bass,
Phys. Rev. C {\bf 68}, 044902 (2003).

\bibitem{coal}
Z.~w.~Lin and C.~M.~Ko,
Phys. Rev. Lett. {\bf 89}, 202302 (2002); R.~J.~Fries, B.~Muller,
C.~Nonaka and S.~A.~Bass,
Phys. Rev. Lett. {\bf 90}, 202303 (2003); D.~Molnar and
S.~A.~Voloshin,
Phys. Rev. Lett. {\bf 91}, 092301 (2003).

\bibitem{phenixv2}
S.~S.~Adler {\it et al.}  [PHENIX Collaboration],
Phys. Rev. Lett. {\bf 91}, 182301 (2003).

\bibitem{sorge97}
H.~Sorge,
Phys. Lett. B {\bf 402}, 251 (1997).

\bibitem{ollitrault92}
J.~Y.~Ollitrault,
Phys. Rev. D {\bf 46}, 229 (1992).

\bibitem{xunu2}
N.~Xu and Z.~b.~Xu,
Nucl. Phys. A {\bf 715}, 587 (2003).

\bibitem{pasi1}
P.~Huovinen, P.~F.~Kolb and U.~W.~Heinz, Nucl.~Phys. A {\bf 698},
475 (2002); P.~Huovinen, P.~F.~Kolb, U.~W.~Heinz, P.~V.~Ruuskanen
and S.~A.~Voloshin,
Phys. Lett. B {\bf 503}, 58 (2001).

\bibitem{pbm}
P.~Braun-Munzinger, K.~Redlich and J.~Stachel,
arXiv:nucl-th/0304013 and references therein.

\bibitem{nxu01}
N.~Xu and M.~Kaneta,
Nucl. Phys. A {\bf 698}, 306 (2002).

\bibitem{zbxu03}
Z.~b.~Xu,
J. Phys. G {\bf 30}, S325 (2004).

\bibitem{starho}
J.~Adams {\it et al.}  [STAR Collaboration],
Phys. Rev. Lett. {\bf 92}, 092301 (2004).

\bibitem{phenixpi0}
S.~S.~Adler {\it et al.}  [PHENIX Collaboration],
Phys. Rev. Lett. {\bf 91}, 172301 (2003).

\bibitem{Ko04}
V.~Greco and C.~M.~Ko,
arXiv:nucl-th/0402020.

\bibitem{xiom130}
J.~Adams {\it et al.}  [STAR Collaboration],
Phys. Rev. Lett. {\bf 92}, 182301 (2004).

\bibitem{Lin:2003jy}
Z.~w.~Lin and D.~Molnar,
Phys. Rev. C {\bf 68}, 044901 (2003).

\bibitem{pythia}
T.~Sj\"ostrand, L.~L\"onnblad and S.~Mrenna,
arXiv:hep-ph/0108264 and references therein.

\bibitem{Greco:2003vf}
V.~Greco, C.~M.~Ko and R.~Rapp,
arXiv:nucl-th/0312100.

\bibitem{starpi0}
I.~J.~Johnson  [STAR Collaboration],
Nucl. Phys. A {\bf 715}, 691 (2003); J.~Adams {\it et al.}  [STAR
Collaboration],
arXiv:nucl-ex/0401008.


\end{thebibliography}
\end{document}